\documentclass{INTERSPEECH2023}
\usepackage{multirow,booktabs,makecell,amsmath,graphicx,amsfonts,color,subfigure}

\interspeechcameraready

\title{Locate and Beamform: Two-dimensional Locating All-neural Beamformer for Multi-channel Speech Separation}
\name{Yanjie Fu$^1$, Meng Ge$^{2,*}$, Honglong Wang$^1$, Nan Li$^1$, Haoran Yin$^1$, \\ Longbiao Wang$^{1,*}$, Gaoyan Zhang$^1$, Jianwu Dang$^1$, Chengyun Deng$^3$, Fei Wang$^3$
\thanks{* denotes the corresponding author.}}
\address{
  $^1$College of Intelligence and Computing, Tianjin University, Tianjin, China\\
  $^2$Department of Electrical and Computer Engineering, National University of Singapore, Singapore \\
  $^3$Beijing Xiaoju Technology Co., Ltd., Beijing, China}
\email{fuyanjie@tju.edu.cn, gemeng@tju.edu.cn, longbiao\_wang@tju.edu.cn}

\begin{document}

\maketitle
 
\begin{abstract}
Recently, stunning improvements on multi-channel speech separation have been achieved by neural beamformers when direction information is available. However, most of them neglect to utilize speaker's 2-dimensional (2D) location cues contained in mixture signal, which limits the performance when two sources come from close directions. In this paper, we propose an end-to-end beamforming network for 2D location guided speech separation merely given mixture signal. It first estimates discriminable direction and 2D location cues, which imply directions the sources come from in multi views of microphones and their 2D coordinates. These cues are then integrated into location-aware neural beamformer, thus allowing accurate reconstruction of two sources' speech signals. Experiments show that our proposed model not only achieves a comprehensive decent improvement compared to baseline systems, but avoids inferior performance on spatial overlapping cases.
\end{abstract}

\vspace{-5px}
\section{Introduction}
\label{sec:intro}
Speech separation aims to extract all individual speech signals from the observed mixture speech, which is a high-demand front-end technology for various real-world applications, such as speech recognition \cite{higuchi2017online, chen2018multi} and speaker verification \cite{rao19_interspeech, xu2021target}. 

For quite some time, multi-channel speech separation solutions have attracted much research attention due to the benefit of spatial information in the microphone array setup. For example, deep learning based beamformers \cite{xu20_interspeech, zhang2021ADL, xu21i_interspeech} are well studied to estimate speech and noise covariance matrices and beamforming weights in an end-to-end fashion. The spatial discrimination facilitates separation when the azimuth difference between speakers is large. Nevertheless, most multi-channel approaches fail on separation when two sources' directions are less than $15^{\circ}$ away from each other. This is the so-called \textit{spatial overlapping} or \textit{spatial ambiguity} issue \cite{8682470}. 

To retain practicality on small azimuth difference cases, some multi-channel separation systems introduce extra speaker cue \cite{vzmolikova2019speakerbeam,9746221} from an enrolled reference utterance, visual cue \cite{gu2020multi} from face tracking and lip movements, or direction cue \cite{xu20_interspeech, zhang2021ADL, xu21i_interspeech} from an additional visual detection system. Such systems require wide-angle camera, speaker embedding model or lip reading network, which might be more resource-consuming. 

In blind source separation applications, above mentioning cues are not available and only observed mixture signal can be leveraged to process. Previous works \cite{8903121, 9414187, 9949050} make efforts to estimate direction-of-arrival (DOA) for each source to guide the estimation of individual sources' speech signals. They rely on the assumption that all microphones can be approximated to have the same DOA when distance between source and microphone array is large. However, this assumption does not hold when source locates in near-field of a large-spacing microphone array. Moreover, obtaining only one DOA is not sufficient for 2D localization. Thus we offer a 2-observer solution rather than using the centroid of array as reference point, which means determining source's two DOAs at two outermost microphones of a linear array.  
This enables us to obtain source's $(x,y)$ coordinates on a 2D horizontal plane through triangulation and leverage these 2D location cues to provide extra discrimination, thus can better handle small azimuth difference cases.

In this work, to explicitly guide separation task with 2D location cues and relieve performance decline problem under spatial overlapping conditions, we propose \textbf{L}ocate \textbf{a}nd \textbf{B}eamform Network (\textbf{LaBNet}), a 2D location-aware neural beamforming framework. Specifically, we design 2D Locator, aiming at estimating source's two DOAs at two outermost microphones as well as corresponding direction embeddings.
2D coordinates of each source are then calculated with the estimated 2 DOAs. In this way, LaBNet can explicitly utilize spectral, spatial, direction and location features with 2D Locator and location-aware neural beamformer to prompt the difference between speakers from all perspectives. 
Experimental results show that both overall speech separation performance and the lower bound on spatial overlapping cases are substantially improved compared to existing state-of-the-art methods.

\begin{figure*}[t] 
\centering 
\includegraphics[width=17cm]{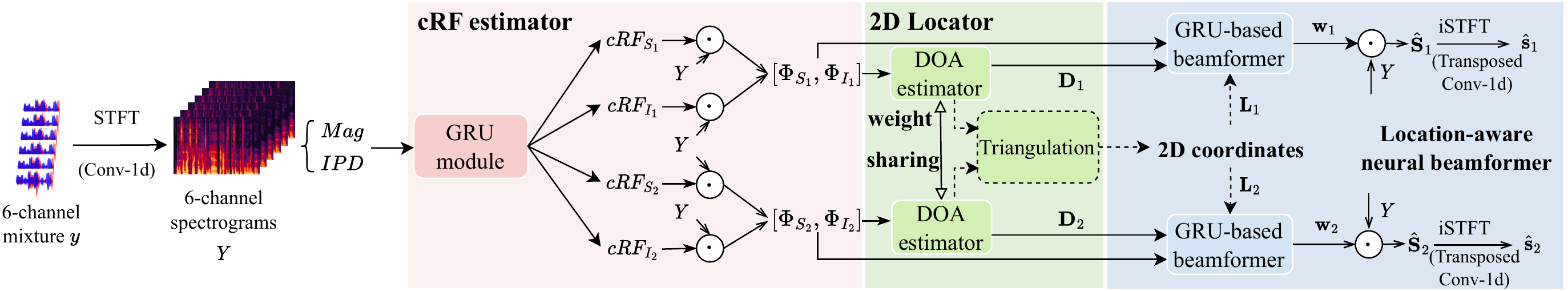} 
\caption{Architecture of our proposed LaBNet. cRF estimator (\textcolor[rgb]{0.9960,0.5058,0.4901}{red part}) produces cRFs to obtain speech and interference covariance matrices for both speakers as input of following modules. 2D Locator (\textcolor[rgb]{0.5961,0.9412,0.5961}{green part}) is introduced for explicit discrimination guidance to neural beamformer with predicted direction and location embeddings. Location-aware neural beamformer (\textcolor[rgb]{0.5058,0.7215,0.9137}{blue part}) predicts beamforming weights for each source. All components in our framework are trainable, which allows backpropagation all around. 
}
\label{Fig.Overview} 
\end{figure*}
\vspace{-5px}
\section{Generalized RNN-beamformer Baseline}
\label{sec:model}

Consider the input mixture waveform $\mathbf{y} \in \mathbb{R}^{L\times M}$, where $L$ is denoted as the number of audio samples and $M$ as the number of audio channels, the corresponding time-frequency representations after STFT can be formulated as
\vspace{-5px}
\begin{equation}
Y(t,f)=S(t,f)+I(t,f)
\end{equation}
\vspace{-2px}
where $S$ stands for the target speech and $I$ represents the sum of interfering speakers' speech and noise (if any).

GRNN-BF proposed in \cite{xu21i_interspeech} contains two modules, complex ratio filter (cRF) \cite{mack2019deep} estimator and RNN beamformer.
The cRFs are predicted by cRF estimator and then used to calculate the target speech and interference covariance matrices. The RNN beamformer accepts the target speech and interference covariance matrices and generates frame-level beamforming weights for the target speaker.

For each T-F (time-frequency) unit of mixture, the estimated cRF applied to its neighbouring units following
\begin{equation}
\begin{split}
\hat{\mathbf{S}}_{\text{cRF}_i}(t, f)=\sum_{\tau_{1}=-K}^{\tau_{1}=K}\sum_{\tau_{2}=-K}^{\tau_{2}=K}\mathrm{cRF}_{\mathbf{S}_i}\left(t,f,\tau_{1},\tau_{2}\right)\\
*\mathbf{Y}\left(t+\tau_{1},f+\tau_{2}\right)
\end{split}
\label{eq:speech}
\end{equation}
where $\hat{\mathbf{S}}_{\text{cRF}_i}(t, f)$ indicates the $i$-th speaker's speech estimated by ${\mathrm{cRF}}_{\mathbf{S}_i}$, and $K$ defines the number of neighboring T-F bins. Likewise, the $i$-th speaker's interference noise $\hat{\mathbf{I}}_{\text{cRF}_i}(t, f)$ can be obtained by ${\mathrm{cRF}}_{\mathbf{I}_i}$ in a similar manner. Then, the frame-wise $i$-th speaker's speech covariance matrix can be calculated with the estimated target speech and its conjugate transpose by a layer normalization \cite{ba2016layer} shown in Eq. (\ref{eq:covmat}). Another layer normalization is applied for computing speaker $i$'s corresponding interference noise covariance matrix $\mathbf{\Phi}_{\mathbf{I}_i}(t, f)$.
\begin{equation}
\mathbf{\Phi}_{\mathbf{S}_i}(t, f) = \text{LayerNorm}(\hat{\mathbf{S}}_{\text{cRF}_i}(t, f) \hat{\mathbf{S}}_{\text{cRF}_i}^{\mathrm{H}}(t, f))
\label{eq:covmat}
\end{equation}
GRNN-BF believes that a better beamformer solution can be directly learned from the speech and noise covariance matrices by neural network. Thus only one unified RNN model is applied to predict the frame-level beamforming weights directly from the covariance matrices, which can be formulated as
\begin{equation}
\mathbf{w}_{i}(t, f) =\mathbf{G R U}\left(\left[\boldsymbol{\Phi}_{\mathbf{S}_i}(t, f), \mathbf{\Phi}_{\mathbf{I}_i}(t, f)\right]\right)
\label{eq:rnnbf}
\end{equation}
Here, $[\cdot,\cdot]$ denotes the concatenation operation of the $i$-th speaker's speech and interference noise covariance matrices. Next, the STFT representation of mixture $Y(t,f)$ is beamformed with the estimated beamforming weights $\mathbf{w}_{i}(t, f)$ to obtain the estimate of the $i$-th target speaker's speech spectrogram $\hat{\mathbf{S}}_{\text{GRNN-BF}_i}(t,f)$, which can be formulated as
\begin{equation}
\hat{\mathbf{S}}_{\text{GRNN-BF}_i}(t, f) =\mathbf{w}_{i}^{\mathrm{H}}(t,f) \mathbf{Y}(t, f)
\label{eq:beamformed}
\end{equation}
Finally, the separated time-domain waveform of the $i$-th speaker's speech $\hat{\mathbf{s}}_i$ can be converted from the beamformed spectrum $\hat{\mathbf{S}}_{\text{GRNN-BF}_i}(t, f)$ using iSTFT. All of the covariance matrices and beamforming weights are complex-valued. The real and imaginary parts of these complex-valued tensors are concatenated throughout the network.

\section{LaBNet Architecture}
GRNN-BF \cite{xu21i_interspeech} requires prior known DOA of target speaker, which is calculated according to the location of target speaker in the video view captured by a wide-angle camera, thus does not suit applications where only observed mixture signal is available. Meanwhile, the DOA is roughly estimated and its DOA estimator cannot be jointly optimized with neural beamformer, which limits the upper bound performance of the system.

In this section, we introduce LaBNet to utilize discriminative 2D direction and location information merely from the given mixture signal (i.e., there is no requirement for providing any extra auxiliary information in our system), thus alleviating the performance degradation caused by spatial overlapping issue \cite{8682470} without sacrificing the overall separation performance.
\subsection{Overview of LaBNet}
Fig.\ref{Fig.Overview} shows the detailed architecture of our proposed LaBNet, which consists of cRF estimator, 2D locator and location-aware neural beamformer. Specifically, the magnitude of reference channel and interaural phase difference (IPD) features of five microphone pairs are extracted from multi-channel mixture STFT. Unlike GRNN-BF \cite{xu21i_interspeech}, the directional feature captured by camera is not available here and we pass magnitude rather than log power spectrum (LPS) along with IPD to cRF estimator. We replace dialated Conv-1D blocks with a GRU module in cRF estimator due to consideration of saving GPU memory usage. After obtaining cRFs for each speaker's speech and interference, multi-channel target speech and corresponding interference can be computed with Eq. (\ref{eq:speech}). Then, speech and interference covariance matrices are calculated with two layer normalization \cite{ba2016layer} shown in Eq. (\ref{eq:covmat}). The next step is to predict beamforming weights from the covariance matrices. This work proposes to introduce a tiny 2D locator (0.9 M) to offer direction and location discrimination for GRU-based beamformer, even on closely located cases, to learn more accurate beamforming weights and exhibit more effective speech separation capability.
\vspace{-5px}
\subsection{2D Locator}
\textbf{DOA estimator}. For each source, the DOA estimator accepts its speech and interference covariance matrices and output the frame-level spatial spectrum(s) \cite{he2018deep} related to its DOA(s). The likelihood-based spatial spectrum is based on the assumption that the probability of source being at each individual angle follows Gaussian distribution that maximizes at ground truth DOA. Formulaically, we adopt a 210-dimensional vector $p_i^{(n)}(\theta)$ for each time frame to include the probabilities of one source locating at 210 individual directions (the azimuth $\theta$ ranges from $-15^{\circ}$ to $195^{\circ}$):
\begin{equation}
p_i^{(n)}(\theta)=e^{-{d\left(\theta,\theta^{\prime}\right)^{2}}/{\sigma^{2}}}
\end{equation}
where $\theta^{\prime}$ is the ground truth DOA of one source, $\sigma$ is a predefined constant that controls the width of the Gaussian function and $d(\cdot,\cdot)$ denotes the angular distance. Note that $n\in\{1,...,N\}$ and when $N$ is set to 1, only one DOA at the centroid of array is estimated; when $N=2$, 2 DOAs w.r.t. 2 outermost microphones will be estimated.
\begin{figure}[t] 
\centering 
\includegraphics[width=8cm]{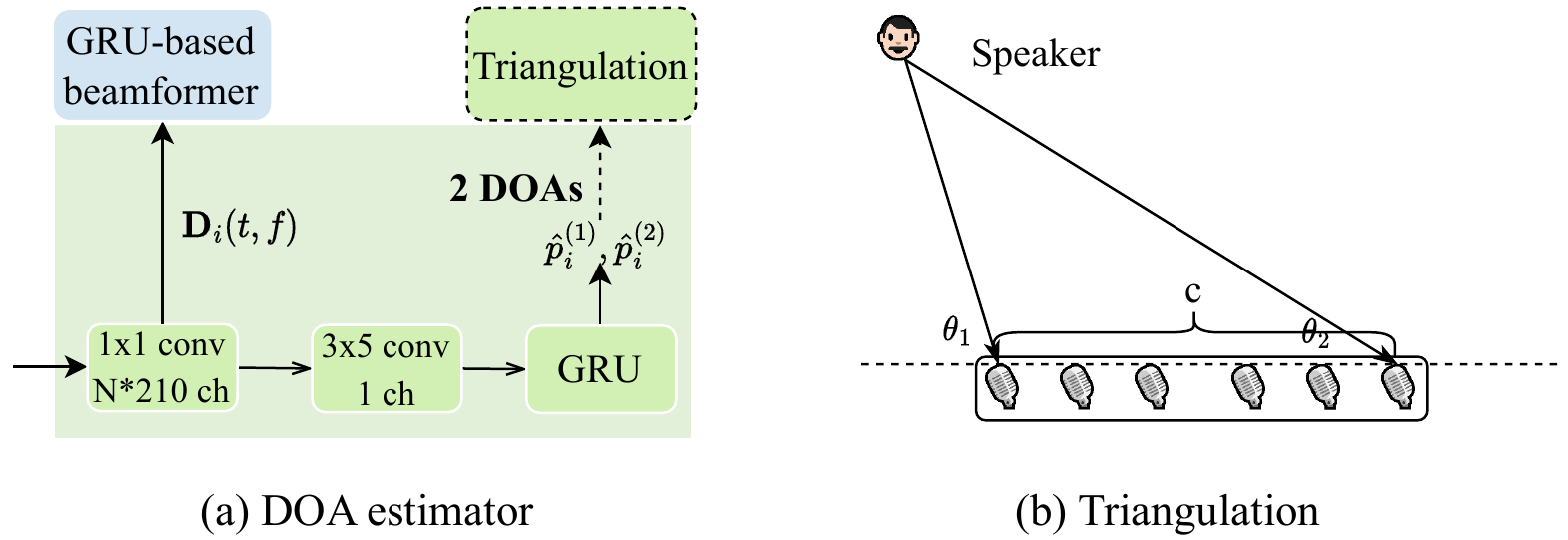} 
\caption{The components of DOA estimator and the diagram of triangulation. Dashed line means optional operation.
}
\label{Fig.Locator} 
\vspace{-20pt}
\end{figure}
The DOA estimator is composed of two convolutional layers and a GRU module. The first convolutional layer projects covariance matrices into the DOA space to obtain $210 \times N$-dimensional direction embedding $\mathbf{D}_i$ for each T-F unit. The second one convolves along time and DOA axes and aggregates features across all frequency bins to generate initial spatial spectrums for each time frame. The following GRU module learns temporal context information among all time frames to polish the frame-level spatial spectrum before outputing it. We select mean squared error to measure the disparity between the estimated and the ideal frame-level spatial spectrum for the $i$-th speaker $\hat{p}_i^{(n)}$ and $p_i^{(n)}$ at $n$-th observer:
\vspace{-5px}
\begin{equation}
\mathcal{L}_{\text{DOA}_i}=\sum_{n=1}^N||\hat{p}_i^{(n)}-p_i^{(n)}||^2
\end{equation}
\vspace{-5px}

\textbf{Triangulation}. Given prior known interval between 2 outermost microphones $c$ and the estimated 2 DOAs regarding them $\theta_1$, $\theta_2$, the distances between source and outside microphones can be given by Law of Sines (Fig.\ref{Fig.Locator} (b)). We then compute 2D coordinates $(x,y)$ which represent source's relative location to the reference microphone. 

\subsection{Location-aware neural beamformer}
Two additional inputs are taken by our proposed location-aware neural beamformer besides speech and noise covariance matrices compared to GRNN-BF: 1) DOA estimator's intermediate output, the direction embedding $\mathbf{D}_i$, that indicates from which azimuth direction the speech comes in each T-F bin. 2) the location embedding $\mathbf{L}_i$, which is the repeats of source's frame-level 2D coordinates $(x,y)_i$ along the frequency dimension, indicating each source's location in each T-F unit.

The estimated direction and location embeddings for each speaker, $\mathbf{D}_i(t, f)$ and $\mathbf{L}_i(t, f)$ are concatenated with speech and interference covariance matrices of corresponding speaker along feature dimension and respectively fed into two parallel beamformers to generate the beamforming weights for each speaker following Eq. (\ref{eq:dbrnnbf}) and thus can more accurately separate the two speaker's speech.
\begin{equation}
\mathbf{w}_{i}(t, f) =\mathbf{G R U}\left(\left[\boldsymbol{\Phi}_{\mathbf{S}_i}(t, f), \mathbf{\Phi}_{\mathbf{I}_i}(t, f), \mathbf{D}_i(t, f), \mathbf{L}_i(t, f)\right]\right)
\label{eq:dbrnnbf}
\end{equation}

In this way, the 2D Locator explicitly performs an auxiliary task for blind source separation. The introduced direction and location embeddings just compensate the less discriminative spatial information in covariance matrices when sources come from adjacent directions, thus enabling the neural beamformer to regain discrimination even on closely located cases.

\subsection{End-to-end Training}
Given the mixture $\mathbf{y}$, target speech $\mathbf{s}_i$, estimated target speech $\mathbf{\hat{s}}_i$, weighted source-to-distortion ratio (wSDR) \cite{choi2018phase} can be computed as follows: $\mathcal{L}_{\text{wSDR}_i}(\mathbf{y}, \mathbf{s}_i, \mathbf{\hat{s}}_i)=\gamma\mathcal{L}_{\text{SDR}}(\mathbf{s}_i, \mathbf{\hat{s}}_i)+(1-\gamma)\mathcal{L}_{\text{SDR}}(\mathbf{y}-\mathbf{s}_i, \mathbf{y}-\mathbf{\hat{s}}_i)$, where $\gamma=||\mathbf{s}_i||^2/(||\mathbf{s}_i||^2+||\mathbf{y}-\mathbf{s}_i||^2)$, $\mathcal{L}_{\text{SDR}}(\mathbf{x},\mathbf{\hat{x}})=-\langle \mathbf{x}, \hat{\mathbf{x}}\rangle/(||\mathbf{x}||||\mathbf{\hat{x}}||)$. The wSDR losses of both speakers are added up as speech separation loss. For optimizing the whole system, we use a multi-task loss as training objective:
\vspace{-5pt}
\begin{equation}
\mathcal{L}_{\text{LaBNet}} = \alpha * \sum_{i=1}^2 \mathcal{L}_{\text{DOA}_i} + \beta * \sum_{i=1}^2 \mathcal{L}_{\text{wSDR}_i}
\end{equation}
where $\alpha$ and $\beta$ are the weights of DOA estimation task and speech separation task, respectively.
We adopt azimuth sorting training strategy \cite{9747141}, sorting all sources' labels in ascending order by DOAs. In this way, permutations of DOAs and speech signals are not mismatched and LaBNet can be clear about with which branch to estimate which source during training. 

\vspace{-2px}

\section{Experiments and Analysis}

\subsection{Dataset}
We follow the simulation procedure in \cite{yin22b_interspeech}. We simulate 6-channel reverberant speech from LibriSpeech corpus \cite{panayotov2015librispeech} with a linear microphone array with spacings of 4 cm, 4 cm, 12 cm, 4 cm, 4 cm. We use pyroomacoustics \cite{scheibler2018pyroomacoustics} to randomly simulate RIRs of 50, 10 and 10 different rooms for train, validation and test dataset split. The sizes (length, width, height) of rooms are ranged from 4m, 3m, 2.5m to 12m, 9m, 5m. The RT60s are sampled between 0.3s and 0.8s. The distance between source and microphone array is set to [0.5m, 8m]. After simulation, we randomly select two source signals of different speakers from the same room to mix and truncate to 4 seconds long segments. The distance between two sources is set to at least 1 meter due to social distancing. Our dataset contains 40,000 utterances (44 hours) for training, 5000 for validation, 3000 for testing. To evaluate ASR performance, we simulate 1000 more utterances which are not truncated and therefore aligned with label text. Examples with azimuth difference of $<15^{\circ}$, $15-45^{\circ}$, $45-90^{\circ}$ and $>90^{\circ}$ account for 16\%, 26\%, 32\% and 26\% in the dataset.
\vspace{-10px}
\begin{table*}[t]
\centering
\small
\caption{Experimental results on our simulated dataset. SI-SDR (dB), PESQ and WER (\%) are evaluted under different azimuth difference ranges between speakers. Bold face indicates the best performance.}
\vspace{-5px}
    \setlength{\tabcolsep}{1mm}{
        \begin{tabular}{|c | c | c c c c c|}
        \hline
        \multirow{2}{*}{Systems\ /\ Metrics} & \multirow{2}{*}{Features \& Setup} & \multicolumn{5}{c|}{SI-SDR (dB) \textbf{$\uparrow$} / PESQ $\uparrow$ / WER (\%) $\downarrow$}  \\
        \cline{3-7} 
        & & $<15^{\circ}$ & $15-45^{\circ}$ & $45-90^{\circ}$ & $>90^{\circ}$ & Avg.
        \\
		\hline
		  Conv-TasNet \cite{luo2019conv} & monaural waveform & 4.35/1.42/72.6 & 4.46/1.42/69.8 & 4.62/1.43/68.8 & 4.75/1.44/68.5 & 4.58/1.43/69.2 \\
		  FaSNet-TAC \cite{luo2020tac} & multi-channel waveform & 3.80/1.34/73.5 & 6.17/1.49/62.4 & 7.80/1.63/49.0 & 8.37/1.68/36.7 & 6.99/1.56/67.1 \\
		\hline
		  NB-LSTM \cite{2022nblstm} & multi-channel STFT & 2.61/1.36/79.1 & 6.56/1.71/57.3 & 9.57/2.09/28.7 & 10.65/2.29/11.9 & 8.13/1.95/32.3 \\
		  GRNN-BF \cite{xu21i_interspeech} & Mag, IPD & 4.94/1.57/46.3 & 8.03/1.94/25.3 & 9.55/2.13/24.4 & 9.92/2.18/13.1 & 8.62/2.02/29.9 \\
		  GRNN-BF-Large & Mag, IPD & 5.18/1.61/45.9 & 8.35/2.01/25.1 & 10.11/2.24/23.0 & 10.55/2.31/12.5 & 9.10/2.11/26.9 \\
		\hline
		  \multirow{3}{*}{LaBNet} & \makecell[c]{Mag, IPD, \\ Estimate 1 DOA} & 6.28/1.78/44.4 & 9.73/2.30/24.1 & 11.42/2.55/18.9 & 11.96/2.68/11.3 & 10.44/2.40/25.2 \\
		\cline{2-7}
		  & \makecell[c]{Mag, IPD, \\ Estimate 2 DOAs} & 6.42/1.79/40.1 & 9.80/2.28/23.3 & 11.53/2.59/16.2 & 11.88/2.64/11.6 & 10.46/2.43/22.6 \\
		\cline{2-7}
		  & \makecell[c]{Mag, IPD, Estimate \\ 2 DOAs \& 2D coordinates} & \textbf{6.62}/\textbf{1.82}/\textbf{39.0} & \textbf{10.01}/\textbf{2.34}/\textbf{21.5} & \textbf{11.73}/\textbf{2.63}/\textbf{15.7} & \textbf{12.11}/\textbf{2.71}/\textbf{10.6} & \textbf{10.69}/\textbf{2.47}/\textbf{20.4} \\
		\hline
        \end{tabular}}
\vspace{-10px}
\label{Tbl.Separation}
\end{table*}
\vspace{-5px}
\subsection{Experimental setup}
In order to train on a single GeForce RTX 3090, we conduct all our experiments with a batch size of 4. When training LaBNet, we use one warm-up epoch and Adam optimizer \cite{kingma2014adam}. The initial learning rate is set to 1e-4 and the max norm of the gradients is set to 3. For the first 10 epochs, LaBNet is trained with emphasis on DOA estimation ($\alpha$ = 5, $\beta$ = 1). And for rest epochs, $\alpha$ and $\beta$ are set to 1 and 10, respectively. For ablation study, we train GRNN-BF with only separation loss after removing 2D Locator of LaBNet. We apply uPIT strategy \cite{kolbaek2017multitalker} when training all systems except LaBNet which is trained with azimuth sorting training strategy. The model achieves the best performance on validation set is chosen after 40 epochs. STFT is conducted with 512-point FFT along 32ms Hamming window with 50\% stride, and the output feature dimension is 257. The STFT and iSTFT are implemented with fixed convolutional encoder and decoder \cite{gu2020multi}. The cosIPDs \cite{wang2018multi} are computed from five pairs between the first microphone and the rest microphones. The cRF estimator is a 2-layer uni-directional GRU with 500 hidden units followed by 4 FC layers with ReLU activation \cite{glorot2011deep}. The GRU module in DOA estimator is a 2-layer uni-directional GRU with 210 hidden units and every neural beamformer contains a FC layer and a 2-layer uni-directional GRU with 300 hidden units followed by a FC layer. The size of cRF $K\times K$ is set to $3\times 3$. The constant $\sigma$ in spatial spectrum coding is set to 8. We use pretrained Conformer-CTC Large to evaluate the ASR performance. See https://huggingface.co/nvidia/stt\_en\_conformer\_ctc\_large.



\vspace{-2px}
\subsection{Results and discussion}
The reverberant speech of each source is taken as reference signal to measure SI-SDR \cite{le2019sdr}, PESQ and WER of the separated speech. DOA estimation performance is also tested with accuracy and mean absolute error for LaBNet. Azimuth sorting training strategy is only applied in training stage as azimuth order is not given when inferencing and all results are the ones computed from the permutation with the highest SI-SDR score.
\vspace{-15px}
\subsubsection{Overall speech separation performance}
As shown in Table \ref{Tbl.Separation}, our proposed LaBNet significantly outperforms previous baselines: NB-LSTM \cite{2022nblstm} and GRNN-BF \cite{xu21i_interspeech} in frequency domain, Conv-TasNet \cite{luo2019conv} and FaSNet-TAC \cite{luo2020tac} in time domain. Compared to GRNN-BF, LaBNet obtains 2.07 dB absolute improvement on SI-SDR (i.e., 10.69 dB vs. 8.62 dB). In metric of PESQ and WER, LaBNet achieves 0.45 and 9.5\% absolute improvements over GRNN-BF.

\textbf{Spatial overlapping issue}.
Although the average performance of all multi-channel models is substantially better than that of single-channel model Conv-TasNet, results of category $< 15^{\circ}$ (azimuth difference between sources less than $15^{\circ}$) reflect the limitation of multi-channel methods \cite{luo2020tac, 2022nblstm} on spatial overlapping cases. LaBNet, however, mitigate this issue with a decent average SI-SDR of 6.62 dB and PESQ of 1.82, which manifests the inclusiveness of our system.

\begin{table}[t]
\centering
\small
\caption{Accuracy (\%) (error $< 5^{\circ}$ as correct) and MAE ($^{\circ}$) results reflecting frame-level DOA estimation performance.}
\vspace{-5px}
    \setlength{\tabcolsep}{1mm}{
        \begin{tabular}{|c| c c c c c|}
        \hline
        \multirow{2}{*}{\textbf{Model}} & \multicolumn{5}{c|}{\textbf{Accuracy (\%) $\uparrow$ / MAE ($^{\circ}$) $\downarrow$}}  \\
        \cline{2-6} 
        & $<15^{\circ}$ & $15-45^{\circ}$ & $45-90^{\circ}$ & $>90^{\circ}$ & Avg.
        \\
		\cline{1-6}
		  DOAnet \cite{8553182} & 59.9/5.2 & 75.3/4.4 & 73.8/6.3 & 61.2/9.6 & 68.9/5.5 \\
		\cline{2-6}
		LaBNet & 88.7/2.7 & 93.1/2.0 & 94.3/2.0 & 88.5/3.0 & 91.7/2.4 \\
		\hline
        \end{tabular}}
\vspace{-20px}
\label{Tbl.Doa}
\end{table}

\vspace{-5px}
\subsubsection{Ablation study}
We conduct ablation study on LaBNet about whether to introduce 2D Locator, the number of DOAs to be estimated and whether to estimate sources' 2D coordinates. Note that GRNN-BF (14.6M) with 2D Locator is equivalent to LaBNet (15.5M). For a fair comparison, we scale GRNN-BF to a comparable model size (15.7M) by adding GRU layers, termed as GRNN-BF-Large.
Additionally estimating 1 and 2 DOAs elevate overall performance to 10.44 dB/2.4/25.2\% and 10.46dB/2.43/22.26\% from 9.1 dB/2.11/26.9\%. Adding the estimation of 2D coordinates brings 0.2 dB further improvement on SI-SDR. Thereby, LaBNet gains 1.6 dB SI-SDR, 0.36 PESQ and 6.5 \% WER absolute improvements over GRNN-BF-Large without increasing model size. This verifies the beneficial guidance from direction and location embeddings to neural beamformer for separation task.

Table \ref{Tbl.Doa} compares the accuracy and MAE results of DOAnet \cite{8553182} and LaBNet. Considering that 5$^{\circ}$ is an admissible error \cite{perotin2019crnn}, the DOA estimation performance of LaBNet is quite outstanding (average angular error of 2.36$^{\circ}$), even on spatial overlapping cases. By separately performing DOA estimation for each individual source rather than for all sources in the mixture once, DOAs are precisely predicted. This confirms that DOA estimation and separation in LaBNet benefit each other.

\begin{figure}[t] 
\centering 
\vspace{-5pt}
\includegraphics[width=8cm]{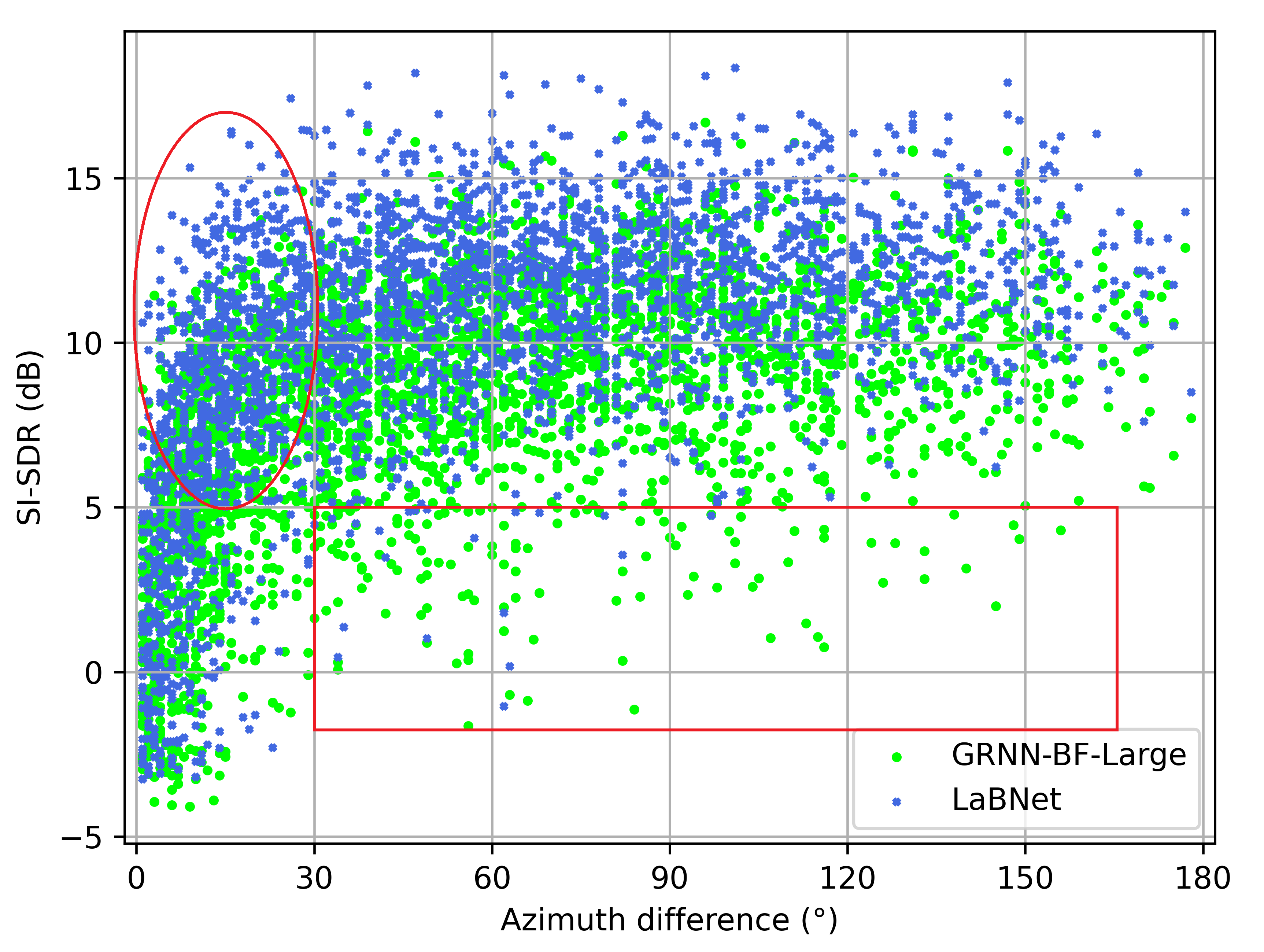} 
\vspace{-20pt}
\caption{The scatter plot of SI-SDR distribution under different azimuth differences between sources. Results of GRNN-BF and LaBNet are reported on testset.
}
\label{Fig.Stats} 
\vspace{-25pt}
\end{figure}


Fig.\ref{Fig.Stats} illustrates distributions of SI-SDR score of GRNN-BF-Large and LaBNet on testset. It can be observed that: (1) In terms of $< 30^{\circ}$cases, LaBNet produces less ``bad cases" with low or negative SI-SDR and elevates more examples to higher than 5 dB SI-SDR (red oval). (2) 2D Location cue does provide extra discrimination for separation, even on large azimuth difference cases, which almost eliminates examples whose SI-SDR is less than 5 dB (red rectangle). This favors backend ASR task a lot. (3) The SI-SDR distribution estimated by LaBNet is more compact with higher mean but lower variance, which suits ASR systems that are sensitive to low SI-SDR cases.
\vspace{-10px}
\section{Conclusion and Future Work}
\vspace{-5px}
In this work, we design a novel multi-channel input and multiple outputs 2D location-aware beamforming network, aiming at providing additional discrimination with direction and location cues. By introducing a light-weight 2D locator (0.9M), our proposed system performs stunningly on both speech separation and DOA estimation tasks. Future work will explore an array geometry agnostic model and consider environmental noise.
\vspace{-10px}
\section{Acknowledgements}
\vspace{-5px}
This work was supported in part by the National Natural Science Foundation of China under Grant 62176182 and DiDi
GAIA Research Collaboration Plan.

\bibliographystyle{IEEEtran}
\bibliography{mybib}

\end{document}